\documentclass[10pt,twocolumn]{article}
\usepackage[margin=0.5in]{geometry}

\usepackage{amsmath}
\usepackage{amssymb}
\usepackage{amsthm}
\usepackage{graphicx}
\usepackage{epstopdf}

\title{Bimodality in the firm size distributions: a kinetic exchange model approach}
\author{Anindya S. Chakrabarti \footnote{Email address: anindya@bu.edu}
\\ Economics  Department, Boston University 
\\ 270 Bay State Road, Boston, MA-02215, USA
 }
\date{\today}

\begin{document}
\maketitle
\begin{abstract}
\noindent Firm growth process in the developing economies is known to produce divergence in their growth path giving rise to bimodality in the size distribution. Similar bimodality has been observed in wealth distribution as well. Here, we introduce a modified kinetic exchange model which can reproduce such features. In particular, we will show numerically that a nonlinear retention rate (or savings propensity) causes this bimodality. This model can accommodate binary trading as well as the whole system-side trading thus making it more suitable to explain the non-standard features of wealth distribution as well as firm size distribution.
\end{abstract}

\section{Introduction}
\label{sec:intro}
\noindent The power-law structure of the firm-size distribution in the developed economies have been investigated thoroughly in the econophysics literature \cite{axtell;01,econphysfirms,lee;98,ishikawa} as well as in economics \cite{simon}. Different classes of models have been proposed that are able to replicate that feature \cite{econphysfirms,podobnik;2008,podobnik;2010,mizuno,riccaboni}. For the developed economies there are two more well established characteristics defining firm growth process viz., Laplace distribution of growth rate and power law decay in fluctuation with size \cite{stanley_et_al;96,stanley;97a,stanley;97b}. Here, we focus on another statistical regularity known for long in the economics literature that in the developing economies the size distribution of firms shows bimodality where most of the firms are either very large or very small \cite{tybout;00}. This particular feature can be dependent on many factors. Here, we will discuss one possible mechanism based on the literature of the kinetic exchange models.
\medskip

\noindent We will introduce a modified kinetic exchange (KE hereafter; \cite{acbkc;07}) model that can replicate this feature in a non-trivial way. The variable of interest is the firm size which is usually measured in terms of its work force. Following Ref. \cite{chakrabarti;12}, we assume that the workers change their workplace every year (the turn-over rates measuring inflow and outflow of workers are very high in the formal sectors). Hence, every firm retains a certain fraction of workers and fires the rest (or the workers themselves decide to leave) giving rise to a permanent cycle of workers across the firms. Though, we borrow the tools of the KE model, clearly we cannot use a binary exchange mechanism here. In particular, we will extend the usual KE model in two directions. First, we will use $n$-ary trading (or collisions, where $n\le N$ the system size) as has been done in Ref.\cite{chakrabarti;12} and secondly, we shall consider the retention rate $\lambda$ to be a function of the evolving variable, the work-force $w$  (known as the {\it savings propensity} and {\it wealth} respectively  in the wealth distribution literature). 

\begin{figure}
\begin{center}
\includegraphics[width=90mm]{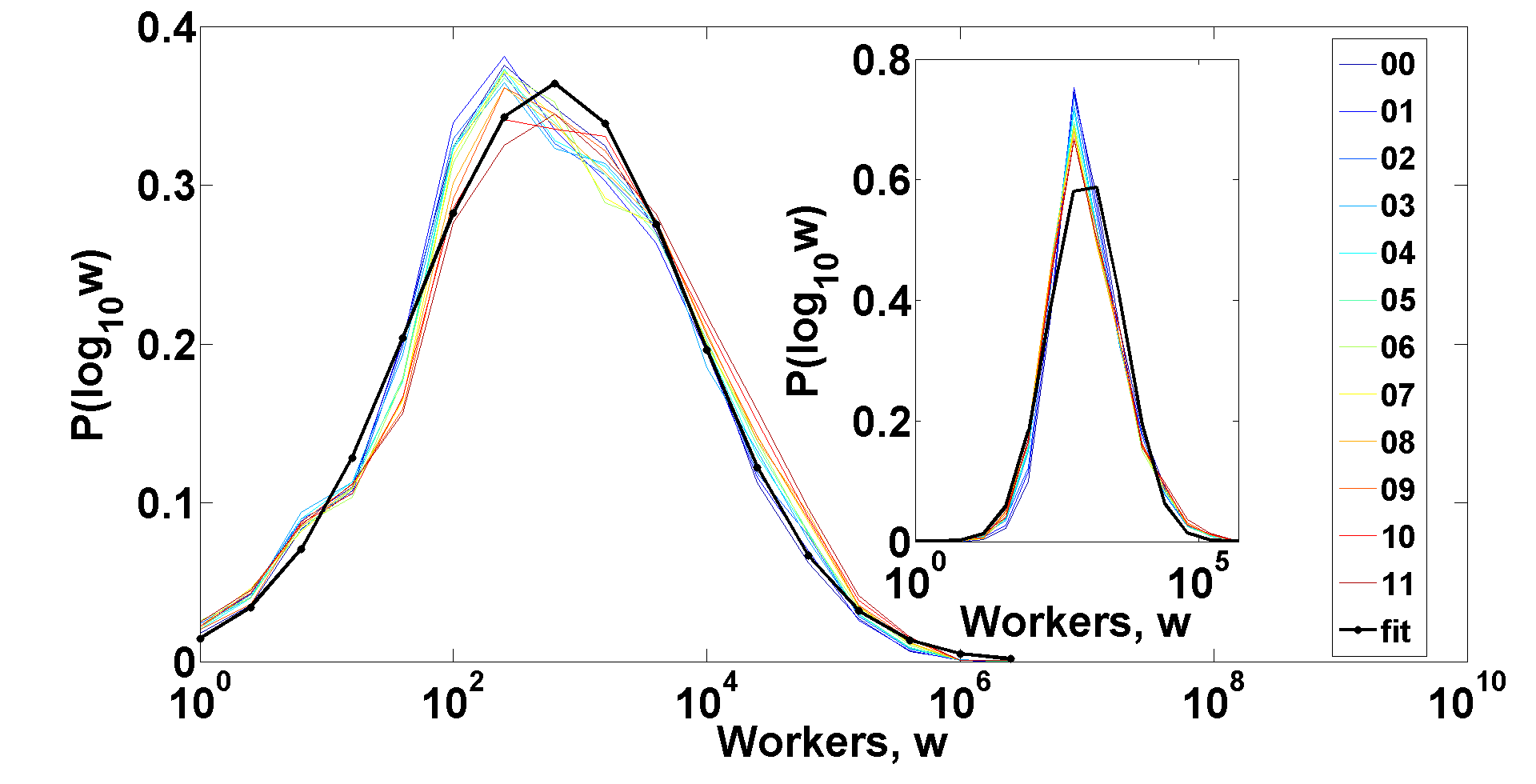}
\caption{ Probability density functions of sizes of all publicly traded firms (2000-2011) in USA and Japan(inset). Unimodality is evident. Data source: Compustat(identifier: GVKEY). Note that this dataset is not exhaustive due to exclusion of publicly untraded firms and hence, it does not show the Zipf's law \cite{axtell;01}. The log of firm-size has a good fit with a normal distribution indicating that the distribution of firm-size is log-normal (color online).}
\label{usa_jpn_firm_size}
\end{center}
\end{figure}

\noindent Here, we present some evidence on such divergence in the growth process of firms in the developing economies. Fig. \ref{usa_jpn_firm_size} shows the distribution of all publicly traded firms in USA and Japan. Clearly, the data shows unimodality. Here, one must be careful because this data-set is limited as it includes only the publicly traded firms (Data source: Compustat; identifier: GVKEY). Since it is not exhaustive, it does not show the power-law tail. BLS \cite{bls} would be more appropriate for that purpose (used in e.g. Ref. \cite{axtell;01}). However, since we are concerned about the uni-modality rather than the exact distribution, this data set suffices. Next, we show that there is a significant pattern in development process where the poorer a country becomes, the more it loses the middle-sized firms. This is known as the `missing middle' \cite{tybout;00}. Here is a genuine problem as such detailed data-bases are not available for the poor economies precisely for which we need data to establish bimodality. One major reason for this lack of data is that most of the firms are unregistered and hence not counted leading to a lower count for micro firms. However, we can use the SME database \cite{SME} which describes the fraction of employees in the micro, small, medium and large firms across the economies (see appendix for data description). The result is shown in the appendix in table \ref{tbl}. Clearly, in the less developed economies the small sized firms are missing \cite{tybout;00}. Below we present the firm-size distributions in four economies (in decreasing level of prosperity ; high, upper-middle (u.m.), lower-middle (l.m.) and low-income resp.) to show the presence of the `missing middle'. The reason for choosing exactly these four economies is that they have identical definitions of firm-sizes. Each entry shows the fraction of the working  population in that economy working in the respective category of firms.

\begin{center}
    \begin{tabular}{ | l | l | l | l |l|}
    \hline     
          &micro  & small & medium & large  \\ \hline
Korea (high)& 33.1 & 17.1 & 36.3 & 13.5 \\ \hline
Latvia (u.m.) & 7.7 &14.1 & 14.8 & 63.4  \\ \hline
Turkey (l.m.) & 38.1 & 8.3 & 17.9 & 35.7 \\  \hline
Kyrgiz Republic (low) & 4.2 & 0 & 0.7 & 95.1      \\ \hline
\end{tabular}
\label{tbl_example}

  \end{center}
\noindent We show that the developing economies are characterized by non-fixed, size-dependent heterogeneity giving rise to bimodality and the developed economies are characterized by fixed heterogeneity giving rise to Zipf's law in the firm size distribution. It is noteworthy that in the literature, bimodality in the wealth distribution is also seen, most prominently in Argentina \cite{ferrero} and in income distribution \cite{lawrence}. We will also show that our model can replicate bimodality even with binary trading mechanism thus explaining the bimodal asset distribution as well. It is noteworthy that there is evidence for bimodality in the world wide income as well (a bunch of poor economies form a mode and the rich economies form another; see e.g. \cite{quah;96}, see also \cite{sala-i-martin;06}) though our model is not directly relevant there.

 \section{The model}
\label{sec:model}
\noindent Consider an array of $N$ firms with size $w_i$ ($i~\le ~N$). Following \cite{stanley_et_al;96} we adopt the rule that the firm-size is determined by the number of their employees. Ref. \cite{chakrabarti;12} first applied the kinetic exchange model to explain firm-size distribution. Here, we modify the retention rate (which corresponds to the savings propensity ($\lambda$) in the usual KE models) as a function of the current labor force $w$. This process captures the more realistic scenario that as a firm increases its work-force, the more workers it retains (or in the context of wealth distribution, a richer person saves more). 
\noindent In general, we suppose that the number of firms from which the workers are leaving and moving into, is $n$. At each time point $(1-\lambda(w))$ fraction of the workforce of those $n$ firms leaves. So there would be a total pool of workers that wants to change their workplace. Next, this pool of workers is randomly divided into those $n$ firms. Hence, the dynamics is given by the following set of equations,
\begin{eqnarray}
w_1(t+1) &=& \lambda(w_1(t)) w_1(t) +\epsilon_1(t+1)\sum^n(1-\lambda(w_j(t))){ w_j(t)} \nonumber\\
\ldots \ldots \nonumber\\
w_i(t+1) &=& \lambda(w_i(t)) w_i(t) +\epsilon_i(t+1)\sum^n(1-\lambda(w_j(t))){w_j(t)} \nonumber\\
\ldots \ldots\nonumber\\
w_n(t+1) &=& \lambda(w_n(t)) w_n(t) +\epsilon_n(t+1)\sum^n(1-\lambda(w_j(t))){ w_j(t)} 
\nonumber
\end{eqnarray}
\noindent such that $\sum^n_j \epsilon_j(t) ~=~1$ for all $t$ (see appendix on how to generate $\epsilon$). As is evident from above, this is a straight generalization of the usual kinetic exchange models (with $n=2$) that has primarily been used to study the income/wealth distribution models (see Ref. \cite{acbkc;07},\cite{ yako-rosser;09}).
\noindent For notational clarity, define $w=[w_1, w_2, \ldots, w_n]'$ at every time period $t$ and consider $n=N$ i.e. the workers change their respective jobs across the whole population of firms . Let us also denote the transition matrix by $T_{\lambda(w)}$. Therefore, the system of equations shown above can be compactly presented as 
\begin{equation}
w(t+1)=T_{\lambda(w)} w(t).
\label{mat_eqn_N}
 \end{equation}
Let us assume the following functional form of $\lambda$, 
\begin{equation}
\lambda(w)=c_1 (1-exp(-c_2w)).
\label{rule1}
\end{equation}  
Simulation shows that the system stabilizes to a certain distribution of $w^*$ which is a fixed point of Eqn. \ref{mat_eqn_N} i.e. $w^*$ solves
\begin{equation}
w^*=T_{\lambda(w^*)}w^*.
\end{equation} 
\noindent Note that with constant $c_1 \in [0,1]$, $c_2\rightarrow \infty$ and binary interactions, the model is the CC model \cite{acbkc;01}. However, for $c_2<<\infty$, the retention rate $\lambda$  becomes a non-linear function of $w$. The analytical solution of this system is not known (even at the limit $c_2\rightarrow \infty$). Hence, we present numerical results only. We considered $N=10^3$ and $\sum^N w_j~=~N$. Given $c_1$ and $c_2$, after simulating the system for $O$(1000) time periods we arrive at a steady state distribution. We take average over $O$(100) of such steady states to arrive at the final distribution.
\subsection{Results}
\label{sec:result}
\begin{figure}
\begin{center}
\includegraphics[width=90mm]{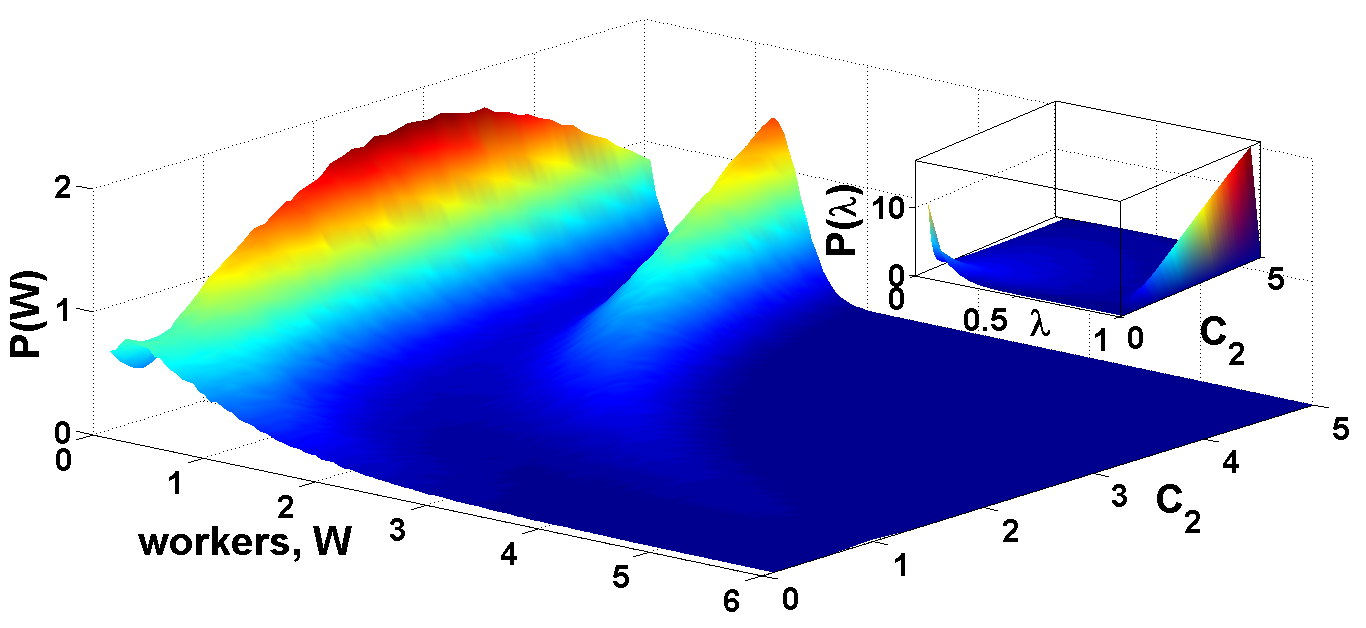}
\caption{ Emergence of bimodality in the basic model with Eqn. \ref{rule1}, is shown with the variation of the parameter $c_2$ with $c_1=0.95$ (color online). In the inset, we show the corresponding distribution of the retention rate $\lambda$.}
\label{bimod_dist_N_exp}
\end{center}
\end{figure}

\begin{figure}
\begin{center}
\includegraphics[width=90mm]{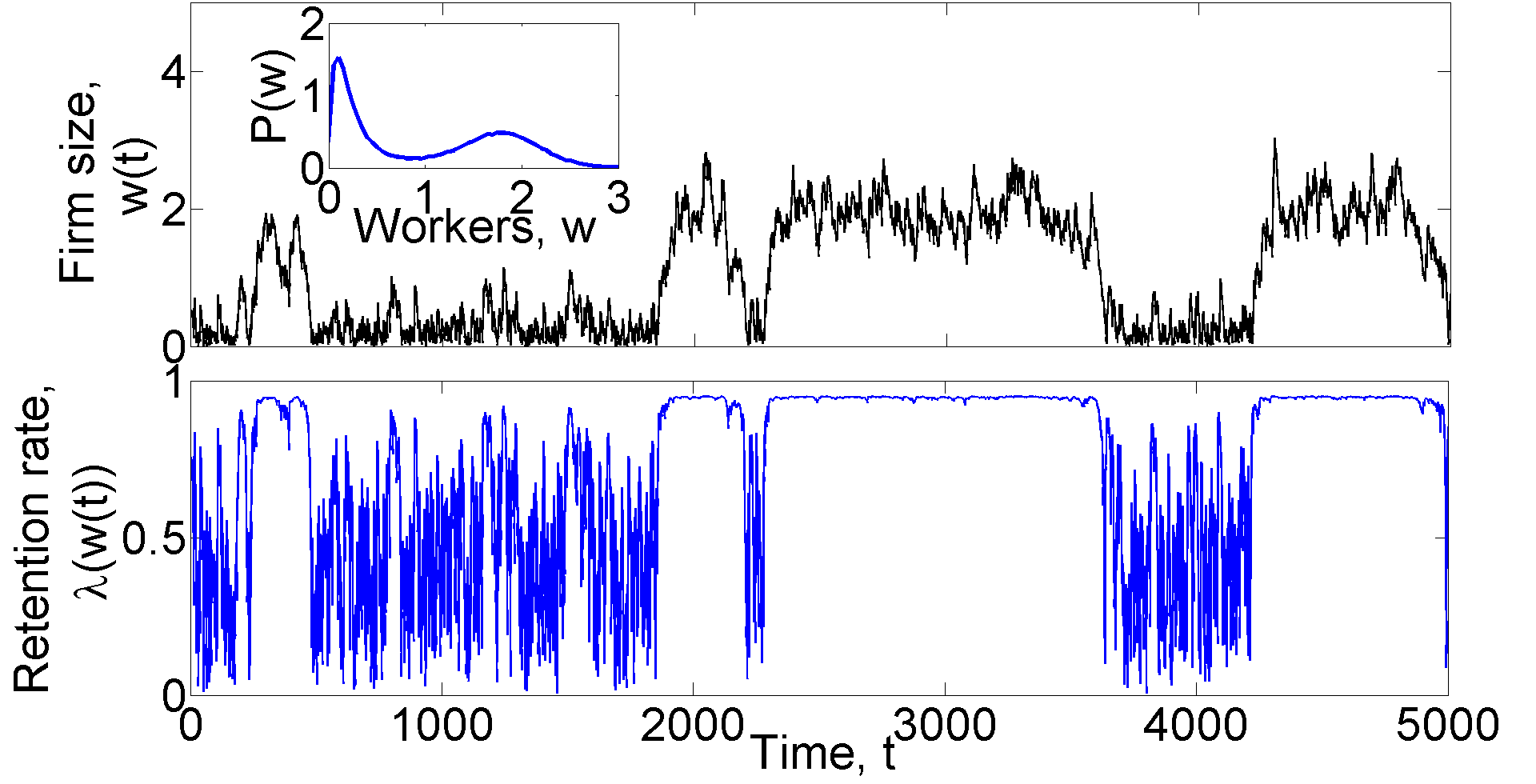}
\caption{ Evolution of size of a particular firm and its retention rate with parameters $c_1=0.95$ and $c_2=3$ (color online). In the inset, we show the distribution of the whole population.}
\label{ts_path}
\end{center}
\end{figure}

\begin{figure}
\begin{center}
\includegraphics[width=90mm]{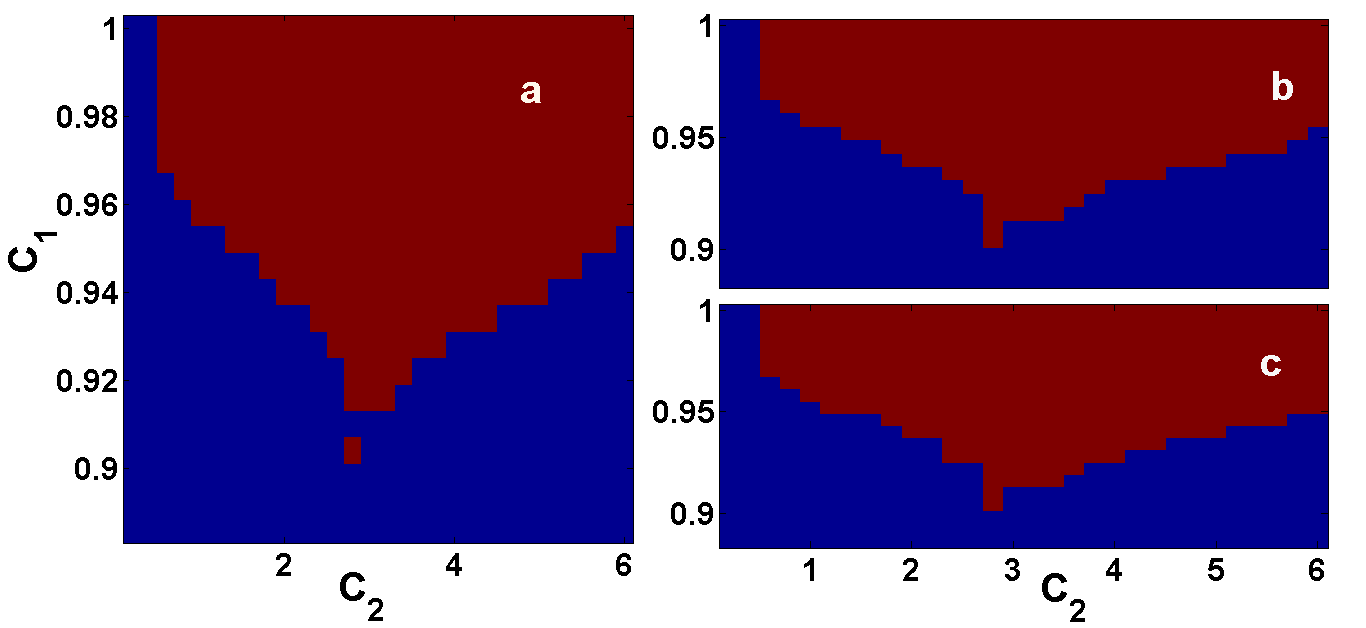}
\caption{ The region in the parameter space($c_1,c_2$) where bimodality appears with Eqn. \ref{rule1} (red squares; color online). For all parameter combinations we have simulated the model and then applied Hartigan's dip test for the presence of bimodality. Panel (a) shows the area where level of significance is 1\%. In panels (b) and (c), the corresponding levels are 5\% and 10\% respectively.}
\label{param_space_exp}
\end{center}
\end{figure}

 \noindent Fig. \ref{bimod_dist_N_exp} shows the steady state distributions for different values of $c_2$ with $c_1$= 0.95. Given the distribution of workers, one can find out the corresponding distribution of the retention rate $\lambda$ as well (see inset of figure \ref{bimod_dist_N_exp}). We also study the evolution of firm-size of a particular firm over time. Fig. \ref{ts_path} shows the time series of firm-size as well as its retention rate moving in tandem. 
\noindent Next, we find out the region in the $c_1, c_2$ space where bimodality appears. For that purpose, we simulate the distribution given a pair of values for $c_1$ and $c_2$. Then we apply Hartigan's dip test \cite{hartigan} to conclude if the distribution is bimodal or not (see Fig. \ref{param_space_exp}). We present the results at three different levels of significance. Note that here the choice is binary, either we accept a distribution to be bimodal or not.  Hence, in Fig. \ref{param_space_exp}, the degree of bimodality is not shown.

\begin{figure}
\begin{center}
\includegraphics[width=90mm]{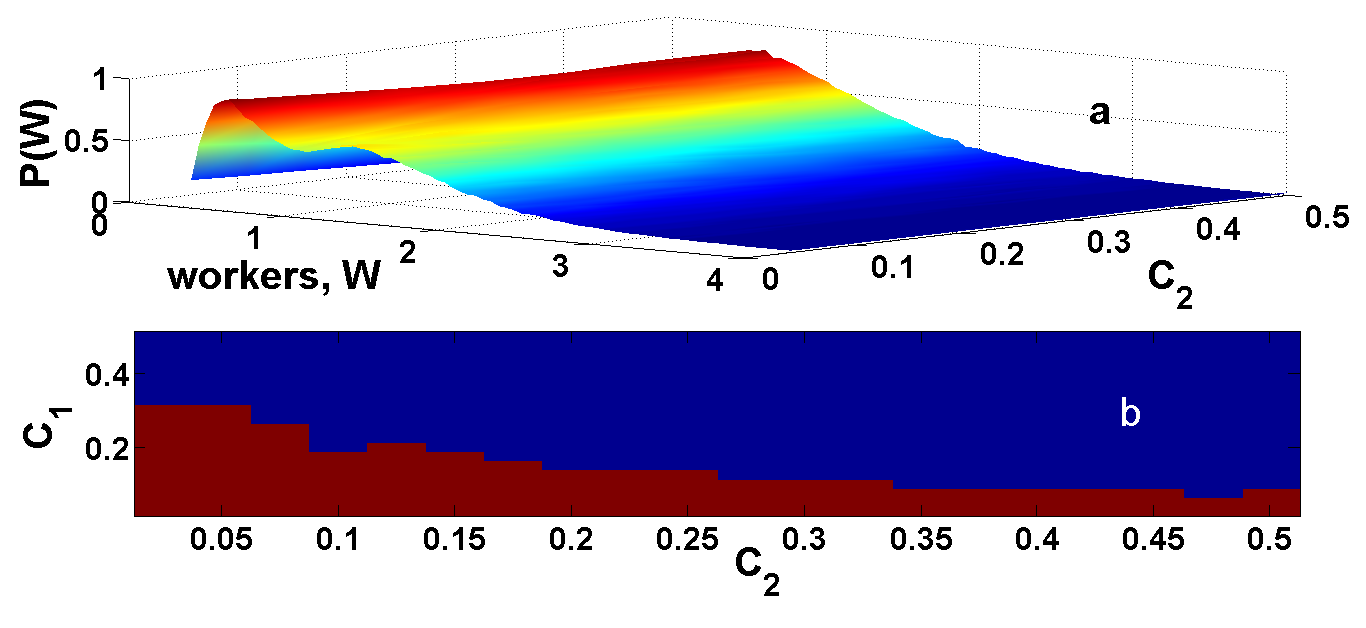}
\caption{Steady state distributions with the sigmoidal retention function given by Eqn. \ref{rule2}. Panel (a):  This graph shows disappearance of bimodality as we vary parameter $c_2$ ($c_1=0.3$). Panel (b):  The region (red squares) in the parameter space where bimodality appears (color online).}
\label{bimod_dist_N_sig_param_space}
\end{center}
\end{figure}

\noindent Ref. \cite{akg;07} shows the existence of bimodality in the KE models by assuming the existence of two separate classes of agents with two widely differing savings propensities ($\lambda$). Here, we adapt our model to include that possibility as well. We considered the following functional form, 
\begin{equation}
\lambda(w)=c_1+(1-2c_1)\left(  \frac{1}{1+exp(-(w-\langle w\rangle)/c_2)}  \right).
\label{rule2}
\end{equation}
\noindent The result is shown in Fig. \ref{bimod_dist_N_sig_param_space}. It is obvious that  $c_1$ $<$ 1/2. The corresponding parameter space where bimodality appears is also shown in the same figure. Note that in the limit $c_2\rightarrow \infty$, this model coincides with the usual CC model with $\lambda$ =0.5 \cite{acbkc;01}. On the other hand, if $c_2\rightarrow 0$, then we get the model proposed in \cite{akg;07}.

\subsection{Transition from purely exponential to Zipf's law}
\label{sec:transition}
\begin{figure}
\begin{center}
\includegraphics[width=90mm]{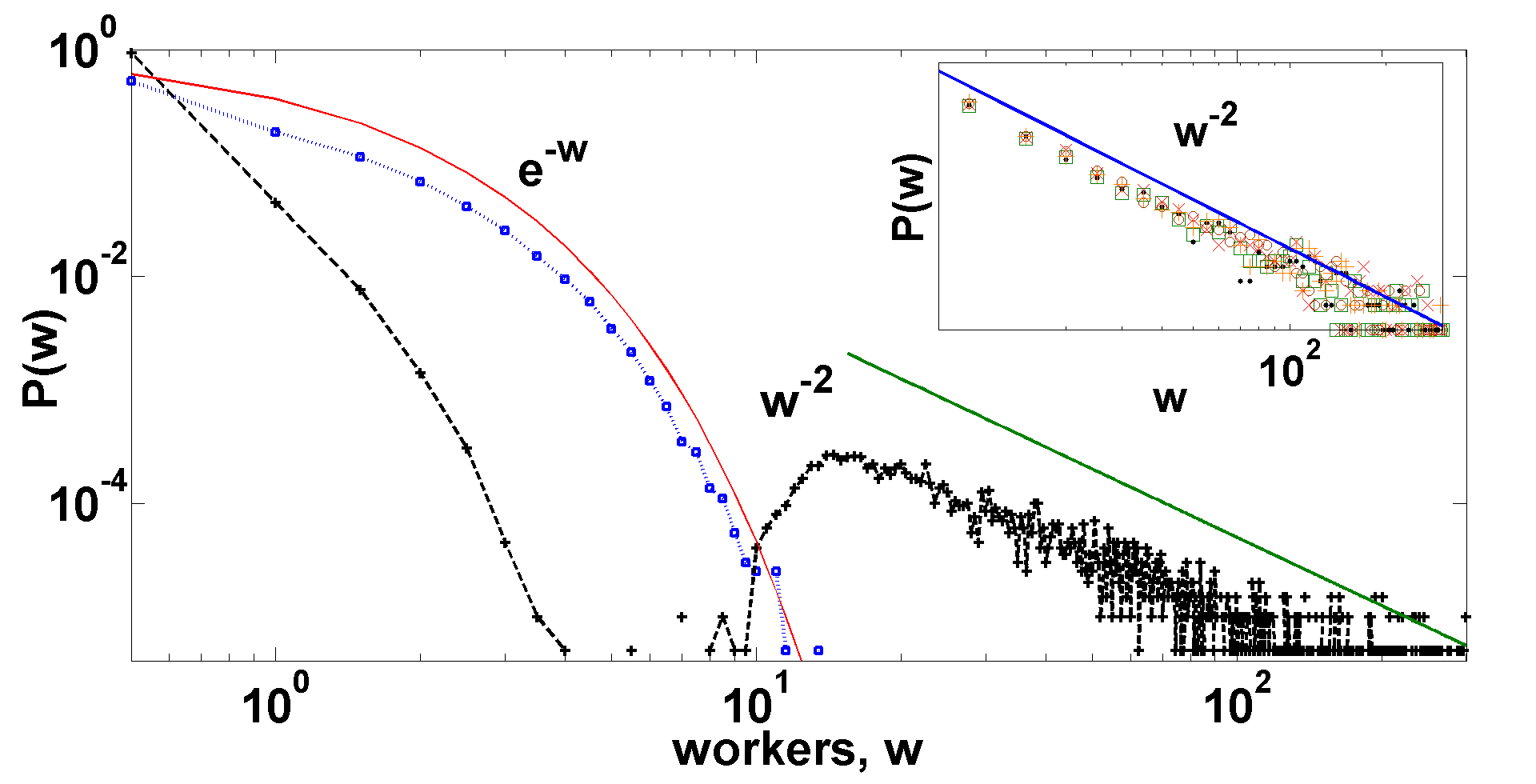}
\caption{This diagram shows the transition from purely exponential to a bimodal distribution with a power law tail due to changes increase in $c_2$ from 0 to 1. The inset shows a complete power law with coefficient -2 for very high values of $c_2$. In all cases, $c_1$ is uniformly distributed for all firms. (color online).}
\label{transition}
\end{center}
\end{figure}

So far we have considered all the firms to be characterized by the same $c_1$ and $c_2$. Hence, they are ex-ante homogeneous but ex-post heterogeneous as their evolution depends on their specific stochastic histories. We can introduce ex-ante heterogeneity as in the CCM model \cite{acbkc;07} where all agents are characterized by different $c_1$ which is fixed over time. With such modification, this model shows a very prominent transition from a purely exponential distribution to a bimodal distribution with a power law tail, eventually converging to a simple power law. We assume $c_1$ to be distributed uniformly between 0 and 1 and fixed for all firms over time. For $c_2$ = 0, the distribution is a purely exponential one as expected (since all $\lambda$s are zero). For $c_2$ = 1, we have a bimodal distribution in the log-log plot with a power law tail (see Fig. \ref{transition}). For higher values of $c_2$, the distribution converges to a power law with coefficient -2. We plotted the data for $c_2$ = 10, 20, 30, 40 and 50 in the inset of Fig. \ref{transition}. 

\subsection{Applications in wealth distribution}
\label{sec:wealth}
\begin{figure}
\begin{center}
\includegraphics[width=90mm]{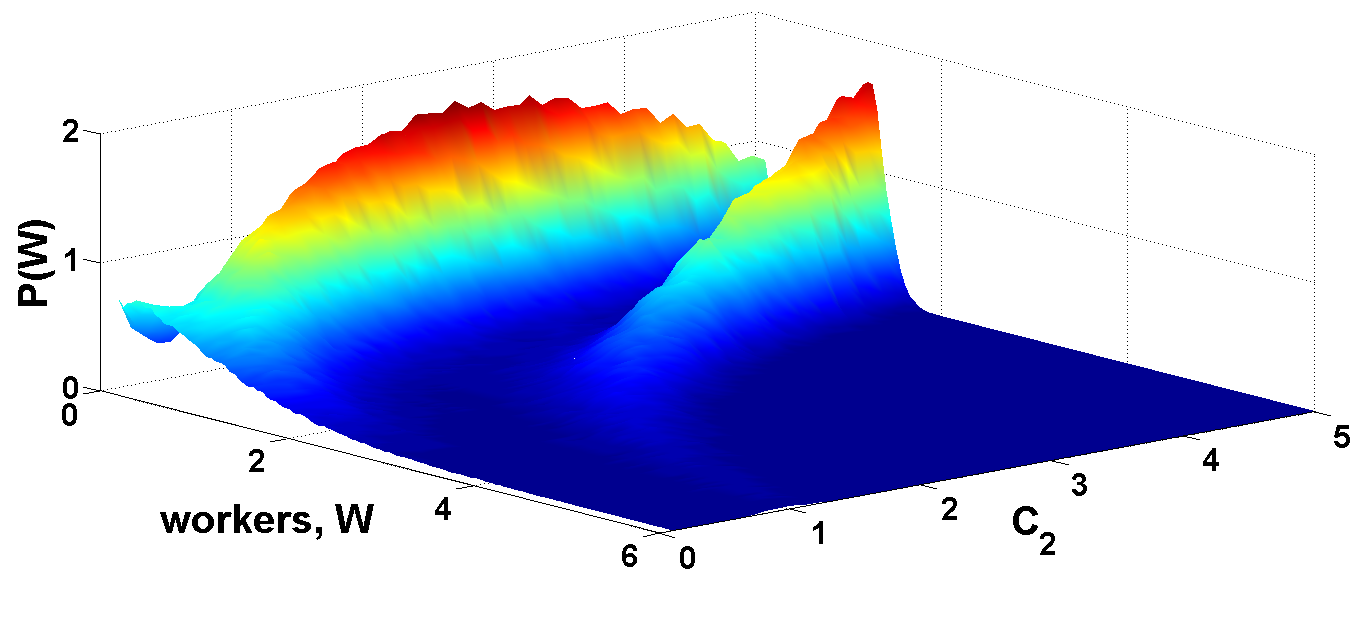}
\caption{Outcome of the model with binary trading mechanism corresponding to the usual kinetic exchange models with $\lambda$ defined by Eqn. \ref{rule1} (color online). Quantitative differences are minor with the case considered with $N$-ary trading (Fig. \ref{bimod_dist_N_exp}).}
\label{bimod_dist_2_exp}
\end{center}
\end{figure}

\noindent It has been claimed in Ref. \cite{ferrero}, that wealth distribution shows bimodality in certain countries. Here, we consider the same model with binary interactions as in the basic kinetic exchange models. All the qualitative results hold true with very little quantitative changes (see Fig. \ref{bimod_dist_2_exp}). However, with $N$-ary interactions, the system reaches steady state much faster than with binary trade.
\section{Zipf's law and Laplace distribution}
\label{sec:zipf}
\noindent Ref. \cite{acbkc;07} derives power law from fixed heterogeneity (assuming a distribution of $\lambda$ fixed for all agents over time). The mean-field equation boils down to
\begin{equation}
\lambda \langle w_{\lambda}\rangle=constant.
\end{equation}
Assuming $\beta(1,n)$ distribution of $\lambda$, $\langle w\rangle$ becomes inverse of a beta distribution which is known to be a power law. However, this model when applied to model firm dynamics \cite{chakrabarti;12} does not show the Laplace distribution of growth. Here, we show that if we consider multiplicative conservation instead of additive (as in the usual KE models), then the Zipf's law and the Laplace distribution in growth are the natural outcomes even {\it without} any heterogeneity.
\noindent Consider the following evolution of $w$,
\begin{equation}
w(t+1)=T_{\lambda(w)} w(t) \hspace{1 cm} \mbox{with $\lambda(w)=0$}.
\label{mat_eqn_N_zl}
 \end{equation}
\noindent To assume complete homogeneity of the firms, define $c_1=0$ in Eqn. \ref{rule1} i.e. $\lambda(w_j)=0~ \forall ~j$. However, lets also redefine measure of workers $m_j=exp(w_j)$. Note that this corresponds to an multiplicative conservation where $\prod_j  m_j= C$ a constant. This equation is readily identified as a production possibility surface of an economy with firms engaged in production sharing the same resources available in finite amounts (this is similar to a consumer's indifference surface).  So with the same amount of resources, the economy can move along the hyper-surface.
 
\noindent We know that 
\begin{equation}
p(w)= T ~exp(-w/T) \hspace{1 cm} \mbox{where $T$ = $\langle$ $w$ $\rangle$.}
\end{equation}
\noindent Assume $T=1$ and recall that $m=exp(w)$. Therefore, 
\begin{eqnarray}
p(m<\theta)&=&\int_0^{log(\theta)} p(w)dw \nonumber \\
  &=& 1-1/\theta
\end{eqnarray}
 \noindent and by differentiation , we get $p(m)\propto m^{-2}$ i.e., we recover Zipf's law \cite{axtell;01}. Note that the growth rate $g_{t+1}=log(m_{t+1}/m_{t})$  (as defined in \cite{stanley_et_al;96}) is the difference of two independent (assuming $\epsilon (t)$ are independent and identically distributed) exponential distributions, 
\begin{equation}
g_{t+1}=w_{t+1}-w_{t}.
 \end{equation}
Note that the characteristic function of $w\sim exp(1)$ is $E(exp(ikw))=1/(1-ik)$ and that of $-w$ is $1/(1+ik)$. Therefore, the characteristic function of $g$ is $1/(1+k^2)$ which is identical to the characteristic function of the distribution, $Laplace(0,1)$. Similar proof goes through even if we consider $T\ne 1$. Hence, the growth rate has Laplace (bi-exponential) distribution \cite{stanley_et_al;96}.

\section{Summary}
In this paper, we have described a kinetic exchange model with a retention rate $\lambda$ (or savings propensity) dependent on the mass of workers $w$ (or wealth) and we have shown numerically that it leads to prominent bimodality in the size distribution of firms as has been empirically found in the developing economies \cite{tybout;00} and in the wealth distribution \cite{ferrero}. To apply our model to firm size distribution, we consider an $N$-ary interaction (see Sec. \ref{sec:model}) whereas to apply it to model wealth distribution, we consider binary interactions with little quantitative change in the results (see Sec. \ref{sec:wealth}). The whole system is described in Eqn. \ref{mat_eqn_N} and \ref{rule1}. As is evident all agents are characterized by the same parameters describing $\lambda$ viz., $c_1$ and $c_2$. The emergence of bimodality as the system organizes itself, is evident in Fig. \ref{bimod_dist_N_exp}. We also show the area in the parameter space where bimodality appears in Fig. \ref{param_space_exp}. To accept or reject bimodality we apply Hartigan's dip test \cite{hartigan}. Ref. \cite{akg;07} considered the possibility of generating bimodality in the basic kinetic exchange model by imposing two widely separated value of $\lambda$. We accommodate that possibility as well by assuming a sigmoidal form of $\lambda$ in Eqn. \ref{rule2}. The results are summarized in Fig. \ref{bimod_dist_N_sig_param_space} as well as the corresponding parameter space where bimodality appears.

Thus in all these cases we assume ex-ante identical agents (that is with identical parametric description) but the agents are ex-post heterogeneous as their evolution depends on their own stochastic paths followed in history. Next, we introduce ex-ante heterogeneity of the agents as in the CCM model \cite{acbkc;07}. We show how the distribution shows a transition from a purely exponential to  bimodality with a power law tail which eventually show a full power law tail with changes in the parameters describing the system. We assumed the same functional form for $\lambda$ as in Eqn. \ref{rule1} but $c_1$ is uniformly distributed (within 0 and 1) over the agents as in the CCM model where $c_1$ differs across agents but are fixed for all agents over time  (see Sec. \ref{sec:transition}). In Fig. \ref{transition}, we show the distributions corresponding to different values of $c_2$. For $c_2$ = 0, the distribution is exponential and as we increase $c_2$ the distribution shows bimodality with the right mode having a power law tail. As we increase $c_2$ even further, the distribution shows a full power law as in the CCM model. In Sec .\ref{sec:wealth} we discuss the case with binary trading and numerically found that the results are almost identical. Lastly, we consider a completely homogeneous agent model and showed that the Zipf's law \cite{axtell;01} and the Laplace distribution \cite{stanley_et_al;96} of fluctuation can be easily accommodated in a slightly different version of the same model.

\medskip

The whole exercise distinguishes our approach to the problem of `missing middle' from other approaches that puts importance solely on either size-dependent or size-independent dynamics. According to our model, the firm-level dynamics is size-dependent or independent depending on the level of development of the economy as a whole. Hence, the firms in the developed economies have fixed-heterogeneity whereas for poorer economies, the firms have size-dependent heterogeneity in the retention rate. To support our theory, we rely on the finding that selection effects are important mostly for the micro firms (which pervades developing economies) and such effects are less prominent for larger firms (most common types in the developed economies) (see \cite{rossi}). This type of selection effects can produce heterogeneity in firm-sizes depending many factors e.g. access to credit market, firm-age, entrepreneurial ability, mobility of inputs etc. Note that most of such features are usually absent (almost by definition) in the poorer economies. There is a huge literature on the effects of financial development (or lack thereof) on the firm dynamics \cite{arellano}. Hence, the firms in the poorer economies have size-dependent dynamics (e.g. a larger firm will have access to credit market whereas a smaller firm may not have any access whatsoever; but in the developed economies all firms have access to credit markets). This justifies our assumption of scale-dependent heterogeneity for firms in poorer economies. However, we abstract from all such details and posit that the heterogeneity is reflected solely in the retention rate which determines the firm's size in our model. This simplification enabled us to economize on the number of variables we study. As has been argued in the literature, bimodality is only a transitional feature seen in the developing economies \cite{tybout;00} or in times of economic instability \cite{ferrero}. We see that ex-post heterogeneity (agents are ex-ante identical but because of the dependence of $\lambda$ on $w$ they are ex-post heterogeneous) induces bimodality. However, as the economy stabilizes the heterogeneity  becomes ex-ante as in the CCM model \cite{acbkc;07} giving rise to a power law distribution.

\section{Appendix}
\subsection{Aggregate firm-size distribution across economies}
Below we present some evidence of the `missing middle' at the aggregate level. The row corresponding to each entry is the income group and the column is the firm-size category. Each entry shows the average fraction of employment in the firm-size category (columns) in the countries in the corresponding income group (rows).

\begin{center}
    \begin{tabular}{ | l | l | l | l |l|}
    \hline     
          &micro  & small & medium & large  \\ \hline
High & 35.6 & 15.6 & 15.4 & 33.4 \\ \hline
Upper middle & 25.8 &13.2 & 19.4 & 41.7  \\ \hline
Lower middle & 29.1 & 10.2 & 16.0 & 44.7 \\  \hline
Low & 11.2 & 7.4 & 9.2 & 72.2      \\ \hline
\end{tabular}
\label{tbl}

  \end{center}

\noindent We have used the data from Ref. \cite{SME} which declares that this data set has secondary data and hence, since sources and definitions differ across countries, there may be errors and there are discrepancies. Therefore, we have removed all the countries with incomplete data and then those countries with absolutely incomparable definitions of micro, small and medium firms. Loosely, we followed the rule that micro firm has less than 10 employees, small and medium have $<$40 or  50 and $<$250 or 300 respectively and the rest are large firms. There are exceptions though e.g. for USA, the middle sized firms have employment between 100 and 499; we incorporated these countries as long as the average employment of one category of  those countries falls within the usual range of that category. We ended up with 22 high income, 3 upper middle, 13 lower middle and 3 low income countries. First, we find out the average employment size in each category (e.g. if a small firm is defined as a firm having workers between 10 to 50, we assume that the average is 30) and we find out the fraction of the employment in a particular category by multiplying the frequency of the firms in each category by that category's average employment. We do this for all countries with (almost) compatible definitions of firm-sizes and then we took average over them to arrive at table \ref{tbl}. We note that micro firms also have a low mass in the low-income group which is probably due to exclusion of a large number of unregistered micro firms. We know that the usage of this data-set is not without its very evident drawbacks. Hence, Ref. \cite{tybout;00} still remains the classic reference on `missing middle'. Another important point is that not all developing economies show a `missing middle' (e.g. Uzbekistan) and similarly, there exists developed economies that shows a `missing middle' (e.g. Denmark).
\subsection{Simulating $\epsilon$}
For two agents, $\epsilon_1$ is uniformly distributed between zero one and we define $\epsilon_2~=~1-\epsilon_1$ . For the general case, note that the problem of generating the vector ($\epsilon_1, \epsilon_2, \ldots, \epsilon_N$ such that $\sum^N \epsilon_j~=~1$ is basically the problem of sampling uniformly from an $N-1$ dimensional simplex. We follow the algorithm described below.
\begin{enumerate}
\item Generate a vector of $N$ variables distributed uniformly within 0 and 1.
\item Take log of them and multiply with -1.
\item  Normalize each of the resulting variables by the sum of them so that their sum is 1.
\end{enumerate}

\end{document}